\documentstyle[prl,aps,twocolumn,epsfig,floats]{revtex}

\begin{document}
\draft
\twocolumn[\hsize\textwidth\columnwidth\hsize\csname %
@twocolumnfalse\endcsname


\title{Subtle Features in Transport Properties: Evidence for a Possible Coexistence of Holes and Electrons in Cuprate Superconductors}
\author{Nie Luo}
\address{Department of Physics and Astronomy, Northwestern University, Evanston, IL 60208}
\date{\today}
\maketitle
\begin{abstract} 

Transport properties of high transition temperature (high $T_c$) cuprate superconductors are investigated within a two-band model. The doping dependent Hall coefficients of La$_{2-x}$Sr$_x$CuO$_4$ (LSCO) and Nd$_{2-x}$Ce$_x$CuO$_4$ (NCCO) are explained by assuming the coexistence of two carriers with opposite charges, loosely speaking electrons ($e$) and holes ($h$). Such a possible electron-hole coexistence (EHC) in other $p$-type cuprates is also inferred from subtle features in the Hall coefficient $R_H$ and thermopower $S$. The EHC possibly relates to the pseudogap and sign reversals of transport coefficients near $T_c$. It also corroborates the electronlike Fermi surface revealed in recent photoemission results. An experimental verification is proposed.
\end{abstract}
\pacs{PACS numbers: 74.25.Fy, 74.25.Jb, 74.62.Dh, 71.35.-y}
]
\narrowtext

Superconductivity in high $T_c$ cuprates is strongly doping dependent \cite{dd}. The parent compounds are antiferromagnetic insulators, which upon proper doping with charge carriers, become superconductive. The type of carriers can be inferred from dopant valences, and is routinely checked by the Hall effect. Although it is natural to assume that a $p$-type sample carries only holes, as is widely believed in the literature, a positive $R_H$ does \emph{not} in theory preclude electrons as the minority carriers. Because of the complicated defect chemistry and band structures \cite{band} in cuprates, one dopant may have bipolar property (i.e., it functions as both donor and acceptor). Before elaborating on this, however, we want to digress to two anomalies, which we think intimately relate to the possible electron-hole coexistence. 

One such anomaly is the normal-state pseudogap. Experiments have revealed various non-Fermi-liquid behaviors in the normal state, particularly the opening of a gap in both spin and charge excitation spectra at a characteristic temperature $T^*$ above $T_c$ \cite{pseudo}. The pseudogap appears in the underdoped regime and weakens as doping level is increased. There is no consensus yet on its origin, but various models have emerged. One approach resorts to the spin-charge separation of Luttinger liquid, originally advocated by Anderson \cite{anderson}. It describes the gap in the spin degree of freedom. However relatively few have been done to the charge sector, where infrared (IR), transport and photoemission \cite{pseudo,batlogg} experiments also reveal gaplike structures. We believe that EHC may help answer this question; electron ($e$) and hole ($h$) attract each other and form excitonic states, resulting in the loss of spectral weight in the charge excitation spectrum.

The other anomaly is the Hall sign reversal near $T_c$ occurring to  properly doped samples. It is often explained in terms of vortex dynamics \cite{tdglwdt}, which explains some phenomena. Recent experiments however have raised serious questions not yet answered by these theories \cite{antivortex}. We argue that vortex motion and pinning are not the whole story because there exists a similar anomaly in the thermopower $S$, usually measured \emph{without} magnetic field. The $S$ anomaly is not widely discussed but it shows up clearly in $S$-$T$ plots \cite{Sanomaly} like Fig. 1(a). It occurs to major cuprate series, on samples from ceramics to single crystals, suggesting a nontrivial nature. This $S$ anomaly is likely a general and doping dependent behavior of cuprate superconductors near $T_c$, just like that in $R_H$. It then poses a problem to theories based on vortices; here we have \emph{no} magnetic field, then where do vortices come? The sign of normal-state $S$ also depends on doping \cite{kaiser}  as shown in Fig. 1(b).  We will see that both are explained naturally in terms of EHC.    
       
\begin{figure} [t]
\begin{center}
\epsfxsize=3.4in 
\epsfysize=1.6in 
\epsffile{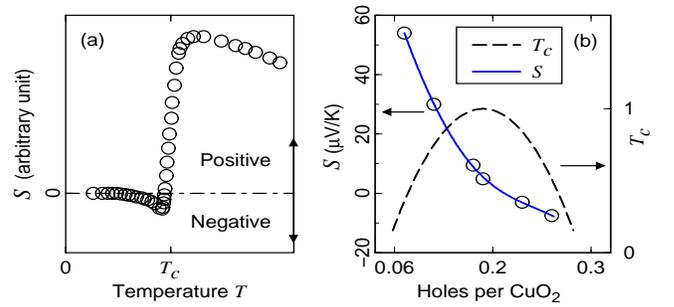}
\end{center}
\caption{(a) Generic thermopower $S$ versus temperature $T$ for properly doped $p$-type samples near $T_c$ [8]. The dot-dashed line is $S=0$. (b) $S$ versus doping for $p$-type cuprates. $S$ values of circles are from Ref. [9] while doping levels are inferred from a general $T_c$ dependence on doping, shown as the dashed line. $T_c$ are normalized to the maximal $T_c$, $T_{cm}$. The solid line is the generic behavior of $S$ in $p$-types although the $S$ values are not to be taken as exact.}
\label{fig1}      
\end{figure} 

The doping dependent $R_H$ in LSCO \cite{ehhall} is shown in Fig. 2(a). The dashed line is the theoretical $R_H$ assuming that each Sr gives one hole. The squares trace the theory relatively well when $x<0.05$. At higher $x$, increasing deviation from the dashed line results in a change in sign. Such a behavior is not explained by a single parabolic band, which requires $R_H\propto1/n$ with $n$ the carrier density. $R_H=0$ means $n \to \infty$, which is clearly unphysical. Microscopic models based on local density approximation (LDA) band theory and the Hubbard model are only partially successful in fitting the curve. The former failed to get an insulator at $x=0$ \cite{band}, while the latter predicted an electronlike Fermi surface (FS) in NCCO \cite{ldashnotgood}, just opposite to experiments.

This crossover, shown as a sharp dip, is however not strange to the two-band, or two-carrier model, which explains reasonably well similar behaviors of $R_H$ as a function of temperature or composition in some chalcogenides like Bi$_2$Te$_3$ \cite{bi2te3}. The term two-band here, should not be taken too literally; it is not needed to have two bands across the FS. The two carriers may simply come from different parts of a single band of complex shape, or just originate from electronlike or holelike portion of the FS. 

\begin{figure} [t]
\begin{center}
\leavevmode
\epsfxsize=3.4in 
\epsfysize=3.2in 
\epsffile{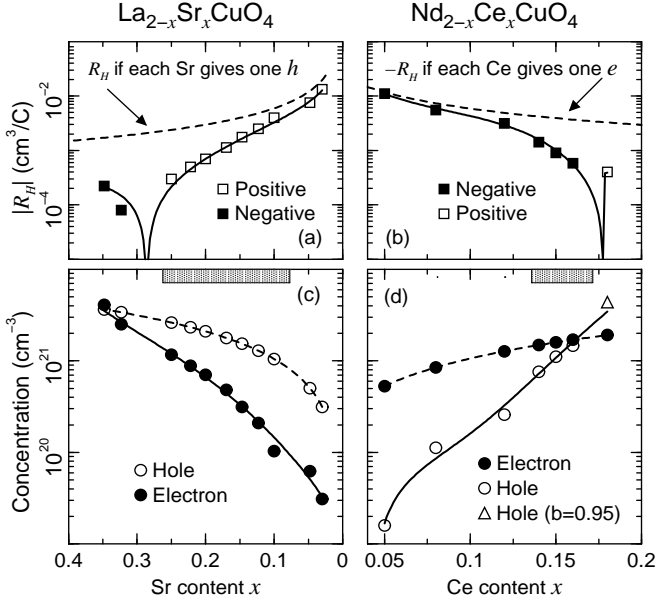}
\end{center}
\caption{$R_H$ in (a) LSCO and (b) NCCO. The squares are after Ref[10], while the dashed theoretical curves assume that each dopant atom gives one $e$ or $h$. Calculated $n_e$, $n_p$ in (c) LSCO and (d) NCCO. Shaded bars indicate superconducting regimes.}
\label{fig2}      
\end{figure}

In the two-band model \cite{2band}, $R_H$ is given by  
\begin{equation}
R_H=\frac{n_p\mu^2_p-n_e\mu^2_e}{e_c(n_p\mu_p+n_e\mu_e)^2},
\end{equation}
where $\mu_p$,  $n_p$ are mobility and concentration of holes respectively, while $\mu_e$, $n_e$ are those of electrons and $e_c$ the elementary charge. In ideal doping model, $n_p=2x/v_u$, where $v_u$ is the unit cell volume and $2$ indicates two formula units per unit cell of LSCO. If there were only holes, $R_H$ would then simply be $v_u/2xe_c$ (i.e., the dashed line). This problem is readily resolved if there are some electrons; a positive $n_e$ in Eq.~(1) makes $R_H$ smaller. Rewrite Eq.~(1) in a compact form, 

\begin{equation}
R_H=\frac{n_p-n_eb^2}{e_c(n_p+n_eb)^2},
\end{equation} where $b=\mu_e/\mu_p$ is the mobility ratio. 

We want to calculate $n_e$ using Eq.~(2). To this end,  we assume: (1) $n_p = 2x/v_u$. (2) $b$ does not vary with doping. The last assumption is crude; $b$ in fact strongly varies with $x$, but we have no better choice because only $R_H$ is known; we have to fix $b$ first in order to get $n_e$.      
Now we estimate $b$. As the zeroth order approximation, we take $R_H$ as coming from majority carriers only and then mobility $\mu=R_H/\rho$ where $\rho$ is the resistivity. This gives $\mu_e$ if $R_H<0$ and $\mu_p$ if $R_H>0$.  We surveyed data in the literature \cite{ehratio,ehncco} from some of the best samples and got $b=1.25$. Details of this survey will be published elsewhere. This $b$ is used for both LSCO and NCCO. Only data at 80 K are taken to reduce errors from $R_H$ variation with $T$. $n_e$ is then directly calculated from the experimental $R_H$. The results are in Fig. 2(c) as solid circles. $n_e$ are then fitted by power series shown as the solid curve that reads $n_e(x)=9.0\times10^{18}+7.3\times10^{20}x+6.3\times10^{22}x^3+2.3\times10^{25}x^{10}$ in cm$^{-3}$. We then plug $n_e(x)$ back in Eq.~(2) to verify $R_H$, resulting in the solid line in Fig. 2(a). Similar result for NCCO is shown in Figs. 2(b) and 2(d) with $n_p(x)=-2.1\times10^{20}+1.0\times10^{21}x^{0.5}+5.3\times10^{26}x^7$ in cm$^{-3}$, which holds if $x\geq 0.05$. (The triangle is a fit using $b=0.95$. The fit cannot be carried out if $b>0.96$. This shows $b$ dependence on $x$.) 

The presence of electron in LSCO is significant. In the superconducting regime, $n_e\sim$ 0.17$-$0.5 $n_p$. As for NCCO, $n_p\sim$ 0.5$-$1.5 $n_e$. Thus EHC effect is better manifested in NCCO, which might explain why it has long been suggested by various groups \cite{ehncco}. Our result this far confirms their works and also gives evidence for EHC in LSCO, a $p$-type superconductor.

One may ask if EHC extends into other $p$-type cuprates where normal-state $R_H$ are positive regardless of doping. However, a sign reversal in $S$ still occurs at high doping levels and near $T_{cm}$, as shown in Fig. 1(b). This strongly suggests an EHC.  Also $R_H$ is somewhat small. The optimal doping is around 0.15 holes per CuO$_2$ (HPC) for most $p$-type cuprates, inferred from the Cu valence obtained from methods like iodometric titration \cite{titration}. However $n_p$ calculated from $1/e_cR_H$ is nearly 2 times as large. The presence of $e$, as in LSCO, is thus suggested. 

We now want to find out $n_e$ but $b$ has to be determined first. $\mu_p$ in these cuprates is much higher than that in LSCO or NCCO, so $b<1$ is likely. However, $\mu_e$ is not known in these materials; and we have to estimate. Because of the possible EHC, transport relaxation now involves multiple processes. There are 3 processes from the inter-carrier (IC) relaxation channel: hole-hole ($h$-$h$), electron-electron ($e$-$e$) and electron-hole ($e$-$h$) scatterings. The rate for $h$-$h$ or $e$-$e$ process is much less than that for $e$-$h$ because of the Pauli principle. Each $e$-$h$ collision involves one $e$ and one $h$ but because $n_p \gg n_e$, each $h$ is scattered less frequently in average. Thus $\mu_p\gg\mu_e$ if only IC channel is counted. However, carrier-defect (CD) scattering also contributes. Defects here are mainly ionized dopants with a density $n_d \approx n_p-n_e$ as required by charge neutrality \cite{phonon}. For electrons, rates from CD and IC are comparable because $n_d \approx n_p-n_e \approx n_p$, while for holes, CD dominates because $n_d \approx n_p-n_e \gg n_e$.  Counting both channels, $\mu_p \approx 2\mu_e$ or $b\approx0.5$ ($b>1$ in LSCO may relate to the K$_2$NiF$_4$ type structure). Also we only take data at $T^*$ because below it many $h$ pairs have already formed, resulting in an underestimate on $n_p$. $T^*$ is selected where $\rho$ begins to deviate below the linear $\rho$-$T$ relation. The results for some of the best samples that we can find in the literature are in Table \ref{table1}. The electron concentration is substantial: $n_e \sim n_p/3.3$ in average, partly confirming results in LSCO. 

\begin{table}
\caption{The $e$-$h$ coexistence in YBa$_2$Cu$_3$O$_{7-\delta}$ (Y-123), Bi$_2$Sr$_2$CaCu$_2$O$_8$ (Bi-2212), Tl$_2$Ba$_2$CuO$_{6+\delta}$ (Tl-2201) and HgBa$_2$Ca$_{n-1}$Cu$_{n}$O$_{2n+2+\delta}$ [Hg-12(n-1)n]. $T_c$ and $T^*$ are in K, $R_H$ in 10$^{-3}$ cm$^3$/C, the assumed $n_p$ and calculated $n_e$ in 10$^{21}$ cm$^{-3}$. $T^*$ is high in Hg-series, which are often underdoped. O: optimal. NO:near-optimal. SC: single crystal. PC: polycrystal. In-plane data for SC samples.}
\begin{tabular}{cccccccc}
Series&$T_c$&$T^*$&$R_H$&HPC&$n_p$&$n_e$&Notes\\
\tableline
Y-123 &93&120&2.33\tablenote{J.P.~Rice, J.~Giapintzakis, D.M.~Ginsberg and J.M.~Mochel, Phys. Rev. B~{\bf 44}, 10158 (1991). Chain contribution removed.}&0.15 &1.73 &0.64&O, SC\\
Bi-2212&82&125&2.60\tablenote{L.~Forro {\em et al.}, Phys. Rev. B~{\bf 42}, 8704 (1990).}&0.15 &1.33 &0.68&NO, SC\\
Tl-2201&85&105&3.42\tablenote{A.W.~Tyler and A.P.~Mackenzie, Physica {\bf 282-287C}, 1185 (1997).}&0.15 &0.87 &0.57&NO, SC\\
Hg-1212&124&320&2.30\tablenote{J.M.~Harris {\em et al.}, Phys. Rev. B~{\bf 50}, 3246 (1994).}&0.15 &1.60 &0.73&NO, PC\\
Hg-1223&135&320&2.38\tablenote{A.~Carrington,{\em et al.}, Physica {\bf 234C}, 1 (1994).}&0.13\tablenote{E.~Pellegrin {\em et al.}, Phys. Rev. B~{\bf 53}, 2767 (1995).} &1.66 &0.64&NO, SC\\
Hg-1234&130&320&2.94\tablenote{J.~Lohle,{\em et al.}, Physica {\bf 223\&224B}, 512 (1996).}&0.13\tablenote{Assumed the same as in Hg-1223}&1.86 &0.21&NO, SC\\
\end{tabular}
\label{table1}
\end{table}

The possible explanation of sign reversals and pseudogap given by EHC justifies our approach this far. While more experiments are needed to establish the $S$ anomaly, it is easily understood in terms of EHC, together with the $R_H$ anomaly in a unified picture. Take $p$-type cuprates for example. Upon cooling, holes form pairs around the pseudogap temperature $T^*$ and then $h$ pairs begins Bose-Einstein condensation (BEC) into supercurrent at $T_c$, resulting into the sharp drop in $\rho$ \cite{becondense}. Because of the large phase fluctuation in these superconductors \cite{phase}, coherence across the whole sample is not yet established near $T_c$ so that the Seebeck or Hall voltage coming from normal carriers is not yet shorted out. 
Because of BEC, $n_p$, the number of normal holes that still contribute to $S$ or $R_H$, is greatly reduced. If it is so reduced that $S_pn_p\mu_p-S_en_e\mu_e<0$, $S$ turns negative. Similarly $n_p\mu_p-n_e\mu_e<0$ results into an $R_H$ anomaly. Whether electrons pair or not does not matter much because $e$ pairs, even if formed, have a lower density and thus enter BEC at a lower temperature $T_c^e < T_c$. Between $T_c$ and $T_c^e$, $e$ pairs are normal and thus still contribute to $R_H$ and $S$ negatively; anomalies are still likely. Our model depends on the delicate balance among $n_e$, $n_p$, $\mu_e$, $\mu_p$, $S_e$ and $S_p$, which explains why it occurs only in properly doped samples. This approach thus emphasizes electronic structures, a point strongly supported by recent experiments \cite{antivortex}. 

$R_H$ anomalies in a few low $T_c$ superconductors such as Nb, V, MoGe/Ge multilayers \cite{socalledanomaly} are often cited as supports for the vortex mechanism. However there exist some confusions. The Hall effect in Nb is strongly impurity dependent. In pure Nb with residue resistivity ratio (RRR) $\gtrsim 2000$ no such anomaly is reported. Reported anomalies in V were all from samples with RRR $\lesssim 200$, which are likely caused by impurities, just as in Nb \cite{hallclarify}.
As for MoGe/Ge, recall that Mo has both electronlike and holelike Fermi sheets \cite{ehsheets}, and Ge is a semiconductor of either $p$- or $n$-type. Thus even the anomalies in low $T_c$ superconductors seem linked more to band structures than to vortex dynamics.     

If the EHC does exist, a question is then asked: where do the electrons come? Oxygen defects like the peroxide ion O$_2^{2-}$ may donate $2e+4h$ besides the $2h$ mode while cation disorders similar to Tl substitution for Ca may also generate electrons \cite{eorigin}. The bipolar doping is alternatively explained in terms of band structures. LDA calculations reveal electronlike and holelike band edges crossing FS for major cuprates \cite{band}, thus making EHC possible. Of particular interest are recent angle-resolved photoemission spectroscopy (ARPES) results on Bi-2212, where electronlike ($e$-like) Fermi surface was found along with the ordinary holelike one \cite{arpes}. Our result strongly corroborates the $e$-like FS from a different perspective and may help to clarify this highly debated issue \cite{edebate}.  
  
The EHC may partly explain the pseudogap. Electrons and holes attract one another to form weakly bound Mott-Wannier (MW) excitons \cite{exciton}. Like that in positronium, the binding energy $E_b = e_c^4M_\mu/32\pi^2\epsilon_0^2\epsilon^2_r\hbar^2$ with $M_\mu$ the reduced mass and $\epsilon_0$ the vacuum permittivity. In Y-123 the carrier ($e$ or $h$) mass $m_c \approx 2m$ \cite{m}, with $m$ the rest mass of a bare electron. The relative dielectric constant $\epsilon_r \approx 14.7$ \cite{dielectric} and thus $E_b \approx 0.063$ eV $\approx 730$ K in $T$. The pseudogap $T^*$ for insulating Y-123 is not known to the author but for LSCO, $T^* \to 720$ K when the metal-insulator boundary is approached \cite{batlogg}. This suggests a good match and thus a possible connection between EHC and the pseudogap. The spectral weight loss in IR at $T$ slightly above $T_c$ might be similarly explained.

As for the stability of electron-hole liquids, we suggest two scenarios. First, $e$ and $h$ are spatially separated, as the result of layered cuprate structure or the stripe phase. Second, most cuprates have electronlike and holelike band-edges well separated in the momentum space, which further reduces the possibility of annihilation.   

\begin{figure} [t]
\begin{center}
\leavevmode
\epsfxsize=2.4in 
\epsfysize=1.6in 
\epsffile{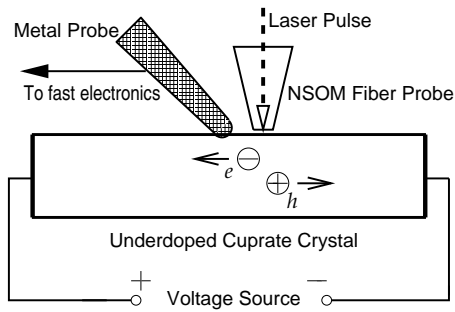}
\end{center}
\caption{Schematic of experimental verification}
\label{fig3}      
\end{figure} 

How this possible EHC relates to the excitonic mechanism of high $T_c$ superconductivity \cite{excitonsuper} is interesting although `excitonic' here is not limited to that of MW or Frenkel excitons. Moreover the question of electronlike FS needs evidence besides that from ARPES. The verification of EHC is thus important and we suggest an experiment to probe the drift direction of carriers under an electric field {\bf E} \cite{drift}. Suppose that a laser pulse is fed through the fiber probe of a near-field scanning optical microscope (NSOM) as shown in Fig. 3. The pulse excites the cuprate sample in a small spot, say 100 nm across, generating carriers, which drift under {\bf E} and upon reaching the metal probe induce a voltage pulse there. By probing the direction at which carriers move relative to {\bf E}, their signs are found. The distance between two probes needs to be small, say $<$ 1 $\mu$m to cope with the fast recombination and small mobility of inequilibrium carriers. The conductance, the doping level of samples, and temperature should be chosen carefully.  

The author is grateful for helpful discussions with Yoshimi Kubo. 

\emph{Note Added.}---J.E.~Hirsch informs the author of similar conclusions from a different perspective. See J.E.~Hirsch and F.~Marsiglio, Phys. Rev. B~{\bf 43}, 424 (1991).

\end{document}